\documentclass[prb, superscriptaddress, showpacs, floatfix,twocolumn]{revtex4}
\usepackage{amsmath}
\usepackage{bbm}
\usepackage{graphicx}
\usepackage{float}
\usepackage{hyperref}
\usepackage{color}
\usepackage{soul}

\begin{document}
%opening
\title{Covalency and the metal-insulator transition in titanate and vanadate perovskites}
\author{Hung T. Dang}
\affiliation{Department of Physics, Columbia University, New York, New York 10027, USA}
\author{Andrew J. Millis}
\affiliation{Department of Physics, Columbia University, New York, New York 10027, USA}
\author{Chris A. Marianetti}
\affiliation{Department of Applied Physics and Applied Mathematics, Columbia University, New York, New York 10027, USA}

\begin{abstract}
A combination of density functional and dynamical mean-field theory is applied to the perovskites SrVO$_3$, LaTiO$_3$ and LaVO$_3$. We show that DFT+DMFT in conjunction with the standard fully localized-limit (FLL) double-counting predicts that LaTiO$_3$ and LaVO$_3$ are metals even though experimentally they are  correlation-driven (``Mott'') insulators. In addition, the FLL double counting implies a splitting between oxygen $p$ and transition metal $d$ levels which differs from experiment. Introducing into the theory an \textit{ad hoc} double counting  correction which reproduces the experimentally measured insulating gap leads also to a $p$-$d$ splitting consistent with experiment if the on-site interaction $U$ is chosen in a relatively narrow range ($\sim 6\pm 1$ eV ). The results indicate that these early transition metal oxides will serve as critical test for the formulation of a general \textit{ab initio} theory of correlated electron metals.
\end{abstract}
\pacs{71.30.+h,71.27.+a}

\maketitle

The Mott insulator\cite{Mott37,Mott1949} is one of the fundamental paradigms of modern condensed matter physics. Many transition metal oxides are believed to be Mott insulators or to undergo ``Mott'' metal-insulator transitions as chemical composition, crystal structure, temperature or pressure are varied, while high transition-temperature superconductivity, high Curie-temperature magnetism, electric-field driven metal-insulator transitions and other important and potentially useful phenomena are believed to be associated with strong electron correlations and the proximity to the Mott transition.\cite{Imada98} Understanding this physics is a crucial goal of condensed matter theory. Recent experimental success in fabricating atomic-precision superlattices involving transition metal oxides with correlated electron properties offers the hope of designing materials with desired correlated electron properties, increasing the need for a predictive theoretical  understanding.\cite{Hwang12}

A complete solution of the all-electron many-body problem for real materials is not available. Many present-day theories proceed by identifying a ``correlated subspace'', a subspace of the full Hilbert space in which electron correlation effects are particularly important and which is treated more precisely while the remainder of the Hilbert space (the background degrees of freedom) is treated by a more computationally efficient mean field-like method. Finally, the solution for the correlated subspace is self-consistently embedded into the background electronic structure.  

For transition metal oxides, the background electrons are typically treated by density functional band theory (DFT).\cite{Jones89} The  correlated subspace is  taken to be the transition metal $d$-orbitals which are defined from the Kohn-Sham eigenstates of the DFT calculation by a  projector \cite{Amadon08,Haule10} or Wannier function \cite{Marzari97,PhysRevB.65.035109} construction. Different DFT formulations for constructing the correlated subspace have been shown to lead to very similar results,\cite{Wang12} provided the localized states are constructed from an  energy range which spans the full transition metal-$d$/oxygen-$p$ complex. The  interactions in the correlated subspace are taken to be the matrix elements of the Coulomb interaction, projected onto the transition metal $d$-orbitals on a given site and screened by other (non-$d$) degrees of freedom.  The interaction matrix elements are either chosen phenomenologically or are calculated using constrained random-phase approximation (cRPA)  \cite{Aryasetiawan04,Aryasetiawan06,Miyake08} or linear response\cite{Cococcioni05} methods. 

The correlation problem is solved using the single-site dynamical mean field approximation \cite{Georges96} and the embedding is accomplished by a combination of the dynamical mean field self-consistency condition, a ``double-counting correction'' \cite{Anisimov91,Czyzyk94,Amadon08,Karolak10} and a charge self-consistency \cite{Kotliar06,Pourovskii07} process. Conceptual and practical uncertainties attend each of these steps, but the resulting density functional plus dynamical mean field theory (DFT+DMFT) approach \cite{Georges04,Held06,Kotliar06} has produced significant insights into the physics of correlated materials. It is therefore important to examine how well the theory does in quantitatively accounting for the properties of real systems. 

In a previous paper \cite{Wang12} focussing on  La$_2$CuO$_4$ and LaNiO$_3$ we found that the double counting physics could usefully be parametrized by $N_d$, the occupancy of the relevant correlated $d$ orbitals. In particular the metal-insulator phase diagram took a nearly universal form when expressed in terms of correlation strength $U$ and $N_d$ while the standard combination of full charge self-consistency and the FLL double counting procedure produced an $N_d$ comparable to that obtained by band theory and predicted that La$_2$CuO$_4$ was deep into the metallic regime of the phase diagram,   although it is normally identified \cite{Anderson87} as a Mott insulator. The paper raised but did not answer the question whether the prediction that La$_2$CuO$_4$ was metallic arose from an incorrect estimation of $N_d$ or a failure of the single-site DMFT approximation.

La$_2$CuO$_4$ and LaNiO$_3$ are ``late'' transition metal oxides in which the transition metal $d$-levels are very close in energy to the oxygen $p$-states; they are identified by Zaanen, Sawatzky and Allen as ``charge transfer'' materials.\cite{Zaanen85} In this paper, we study the early transition metal oxides  SrVO$_3$, LaTiO$_3$  and LaVO$_3$. In these materials the $d$ and $p$ states are well separated in energy and they are normally identified as being in the Mott-Hubbard regime of the Zaanen-Sawatsky-Allen phase diagram. 

\begin{figure}[t]
    \centering
    \includegraphics[width=\columnwidth]{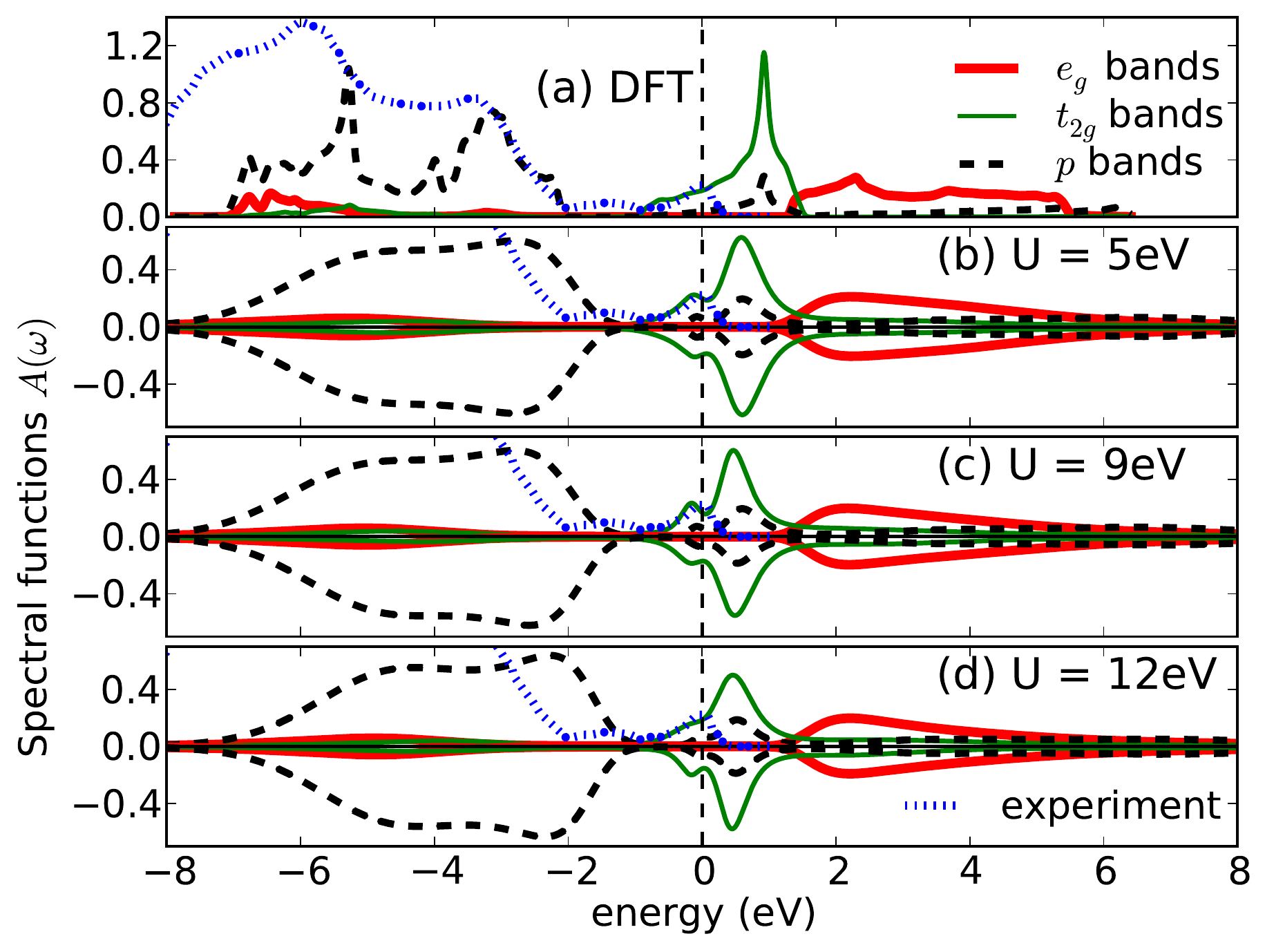}
    
    \caption{\label{fig:spec_fullcharge_SVO_1}(Color online) (a) Density of states from DFT calculation of SrVO$_3$. 
(b,c,d) Positive half-plane: Spectral functions for the $e_g$, $t_{2g}$ and oxygen $p$ bands of SrVO$_3$ obtained from fully charge self-consistent FLL-double counting DFT+DMFT calculations. Negative half-plane: spectral functions obtained from ``one-shot'' calculations \emph{without} full charge self consistency but with $d$ level energies  adjusted so that the $d$-occupancies are within $0.01$ of the values  found in the corresponding fully charge self consistent calculations. The dotted curve (blue online) is the experimental spectra reproduced from Ref.~\onlinecite{Yoshimatsu10}. }
\end{figure}

We begin our studies with SrVO$_3$, which forms in the cubic perovskite structure with nominal V-configuration $d^1$ (one electron in the $d$ shell) and  is experimentally found to be a moderately correlated paramagnetic metal.\cite{Onoda91} The positive $y$ axis portions of the three panels of Fig.~\ref{fig:spec_fullcharge_SVO_1} present electron spectral functions (many body density of states) obtained from  fully charge self-consistent DFT+DMFT calculations of SrVO$_3$ for different values of the $d$-level charging energy $U$ but with the Hund's coupling $J$, which is only very weakly renormalized by solid state effects,\cite{Aryasetiawan04,Aryasetiawan06,Miyake08}  fixed at the value $J=0.65eV$ generally accepted for early transition metal oxides.\cite{Vaugier12} The calculations were performed with the Wien2K+TRIQS code \cite{Aichhorn11,triqs_project} with the full 5-fold degenerate $d$-manifold included and using the FLL double counting \cite{Czyzyk94} at inverse temperature $\beta=10eV^{-1}$. In terms of the Slater parameters $F_0,~F_2,~F_4$,\cite{Czyzyk94} $U$ and $J$ in Kanamori's notations are written as $U = F_0 + 4\left(F_2+F_4\right)/49$ and $J = 5\left(F_2+F_4\right)/98$. To obtain the spectra we used  maximum entropy techniques \cite{Jarrell96} to analytically continue the Green function. In these calculations spin-flip and pair-hopping terms in the interaction are neglected. We have verified \cite{Dang13c} via an extensive study of one parameter set ($U=5eV$, $J=0.65eV$) that the full rotationally invariant and Ising methods give essentially identical results for the early transition metal oxide ($d^1$ and $d^2$ formal valence) cases studied here.

Two key results are evident from Fig.~\ref{fig:spec_fullcharge_SVO_1}. The density of states at the Fermi level is non-zero for all three $U$-values considered within the standard DFT+DMFT scheme: increasing the interaction strength even to very large values does not drive a metal-insulator transition. Also, the spectra for all three $U$-values are very similar; in particular the relative positions of the $p$ and $d$ bands are almost independent of the correlation strength. The fully charge self consistent calculation places the oxygen bands $\sim 1eV$ closer to the $p$ bands than is found in experiment. In the calculations reported in Fig.~\ref{fig:spec_fullcharge_SVO_1} the $d$-occupancies (here defined for the full 5-fold degenerate $d$ manifold) are $N_d^{tot}=2.60$ (DFT with $U,~ J=0$), the fully-charge self consistent $N_d^{tot}$ are $2.55$, $2.51$ and $2.50$ for $U=5,9,12eV$ respectively.  ``One-shot'' calculations with the same FLL double counting yield $N_d^{tot}$ are $2.48, 2.36$ and $2.30$ correspondingly. Thus  as previously found for pnictides\cite{Aichhorn11} and La$_2$CuO$_4$,\cite{Wang12} within this theoretical framework the effects of increasing the intra-$d$ interaction strength are to a very considerable extent compensated by the combination of the charge self consistency and the double counting correction with the full charge self consistency playing an important role. The $N_d$ is  slightly decreased from the band theory value and is weakly $U$-dependent, while  the  $p$-$d$ splitting is essentially independent of $U$. We have performed similar calculations for LaTiO$_3$ and LaVO$_3$ (not shown) in hypothetical cubic and experimental (GdFeO$_3$-distorted) structures, with the same result -- one finds always a metallic state, with the $p$ bands too close to the $d$ bands and an $N_d$ close to the band theory value.

The fully charge self-consistent calculations involve substantial computational complexity in obtaining convergence and  do not permit easy exploration of modifications of the electronic structure that would change the relative positions of the $p$ and $d$ manifolds. However, the negative $y$ axis portions of Fig.~\ref{fig:spec_fullcharge_SVO_1} show that a simpler and more flexible procedure can be used. These panels report the results of  a ``one-shot'' procedure in which the DFT band structure is used to define the $d$-orbitals and then without further charge self-consistency the position of the $d$-level is adjusted so that the DMFT calculation reproduces the $N_d$ found in the fully charge self consistent calculations. The high degree of similarity  of the two sets of spectra shows that the only important role played by the DMFT and charge self-consistency steps is the adjustment of the $d$ occupancy. In the rest of this paper we therefore present one-shot calculations in which the $d$-states are defined from a DFT calculation and the $d$-level energy is adjusted to produce the desired $N_d$ or other physical behavior.

The calculations  presented in the rest of this paper use the pseudopotential-based Quantum Espresso code \cite{QE-2009,QEPseudo}  to obtain energy bands and Wannier methods  as implemented in Wannier90\cite{Mostofi2008685}  to define the orbitals in the correlated subspace. For GdFeO$_3$ distorted structures we choose for each unit cell a local basis aligned to the transition metal-oxygen bond direction. This minimizes off diagonal terms in the dynamical mean field hybridization function and reduces the severity of the sign problem in the CT-QMC impurity solver. Because of the low $d$-valence we include as correlated states in the impurity model only the $t_{2g}$ portion of the $d$ manifold, neglecting the $e_g$ orbitals. We have verified \cite{Dang13c} that  for  both $d^1$ and $d^2$ formal valences this approximation does not change the results. The intra-$d$ interactions which define the correlation problem then take the standard 3-orbital Slater-Kanamori form \cite{Kanamori1963}
\begin{equation}\label{eq:onsite_SlaterKanamori}
\begin{split}
H_{onsite} & = U\sum_{\alpha}n_{\alpha\uparrow}n_{\alpha\downarrow}  + (U-2J)\sum_{\alpha\neq\beta} n_{\alpha\uparrow}n_{\beta\downarrow} + \\
& + (U-3J)\sum_{\alpha > \beta,\sigma}n_{\alpha\sigma}n_{\beta\sigma} + \\
& + J\sum_{\alpha\neq\beta} ( c^\dagger_{\alpha\uparrow}c^\dagger_{\beta\downarrow}c_{\alpha\downarrow}c_{\beta\uparrow}
  + c^\dagger_{\alpha\uparrow}c^\dagger_{\alpha\downarrow}c_{\beta\downarrow}c_{\beta\uparrow}).
\end{split}
\end{equation} 
where $\alpha,\beta$ labels orbitals in the transition metal $t_{2g}$ manifold on a given site.  As in the fully charge self-consistent calculations we neglect the spin-flip and pair-hopping terms in Eq.~\ref{eq:onsite_SlaterKanamori} which speeds up the calculations by factors $\sim 5$ and enables the surveys we present of the phase diagram.  We work at inverse temperature $\beta=10eV^{-1}$ which is high enough for rapid calculation but low enough to reveal the important behavior. The spectra are obtained by using a maximum entropy continuation of the self energy to compute the lattice Green's function. 
We introduce $\Delta$, the double counting correction, which enters the calculations by adjusting the noninteracting $d$ energy level, as $\epsilon_d\to \epsilon_d - \Delta$. $\Delta$ is varied in a wide range and parametrized by the $t_{2g}$-shell occupancy $N_d$.

\begin{figure}[t]
 \centering
 \includegraphics[width=0.9\columnwidth]{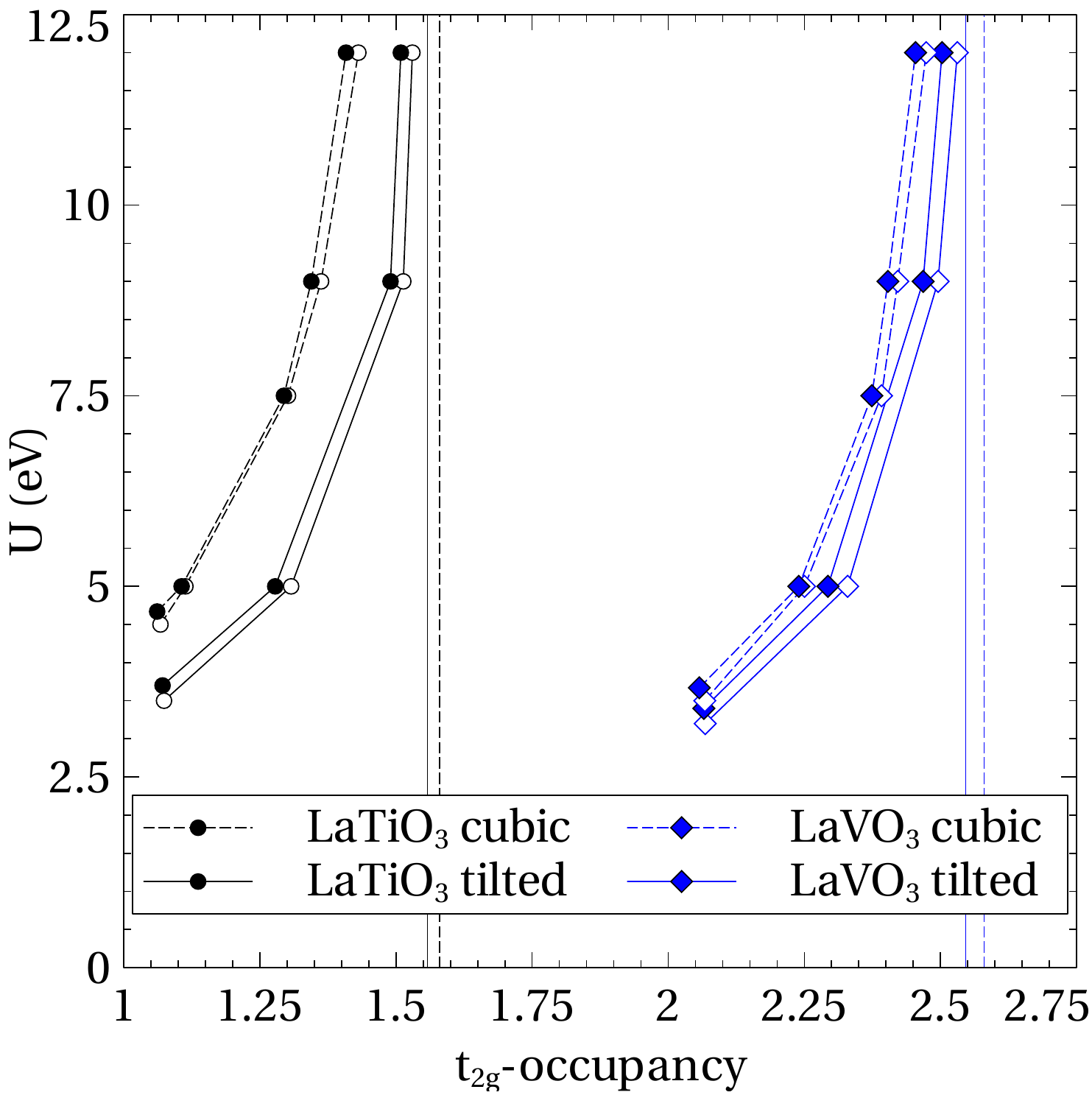}
\caption{\label{fig:dmft_tilted_lto_lvo_1}(Color online) The MIT phase diagrams for LaTiO$_3$ and LaVO$_3$. The solid (dashed) lines are phase boundary and DFT $N_d$ for tilted (cubic) structure. Open symbols represent metallic solutions, closed symbols are insulating solutions.}
\end{figure}

We have calculated properties of LaTiO$_3$ and LaVO$_3$ (experimentally both Mott insulators, with formal $d$ occupancy $d^1$ and $d^2$ respectively\cite{Pavarini04,Raychaudhury07}). We have studied both the experimental (GdFeO$_3$-distorted) structure and  the ideal cubic perovskite structure (with the same mean La-La distance as the experimental structures).
Fig.~\ref{fig:dmft_tilted_lto_lvo_1} shows the metal-insulator phase diagrams of these materials in the plane of interaction strength $U$ and the $t_{2g}$ occupancy $N_d$.  We located the metal-insulator transition from the lattice Green function, computed from the continued self energy as described above.   We identify materials as insulating if the imaginary part of the local Green's function vanishes at the chemical potential. We define the gap magnitude from linear extrapolation of the density of states and identify the metal-insulator transition as the point at which the gap is closed. For reference we show as vertical lines the  $N_d$ predicted by density functional band calculations and we performed extensive standard fully-charge self-consistent calculations and found that the value of $N_d$ is very close to the DFT value.

We see that (as found previously in studies which focussed only on the $t_{2g}$-like antibonding bands \cite{Pavarini04,Raychaudhury07}) the orbital splitting induced by the GdFeO$_3$ distortion has a crucial effect on the metal insulator transition in the $d^1$ titanate and a modest effect in the $d^2$ vanadate. But more importantly,  we see that for both the hypothetical cubic and actual GdFeO$_3$-distorted structures, reasonable interaction strengths $\lesssim 7eV$ locate the metal-insulator phase boundary very far from the $N_d$ predicted by band theory, implying that within the FLL double-counting scheme, standard DFT+DMFT approach predicts the materials to be metallic.

\begin{figure}[t]
    \centering
    \includegraphics[width=0.95\columnwidth]{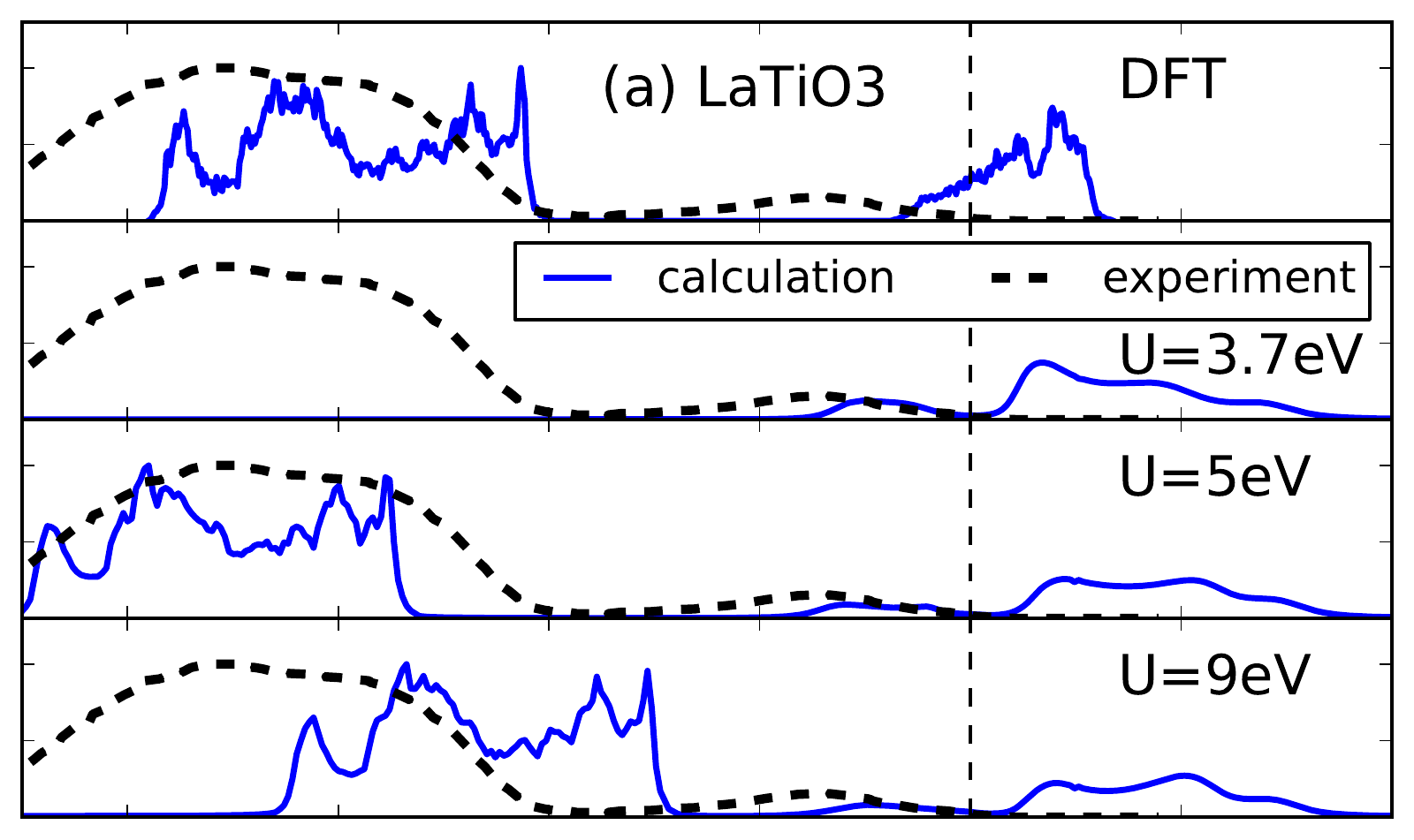}\\
    \includegraphics[width=0.95\columnwidth]{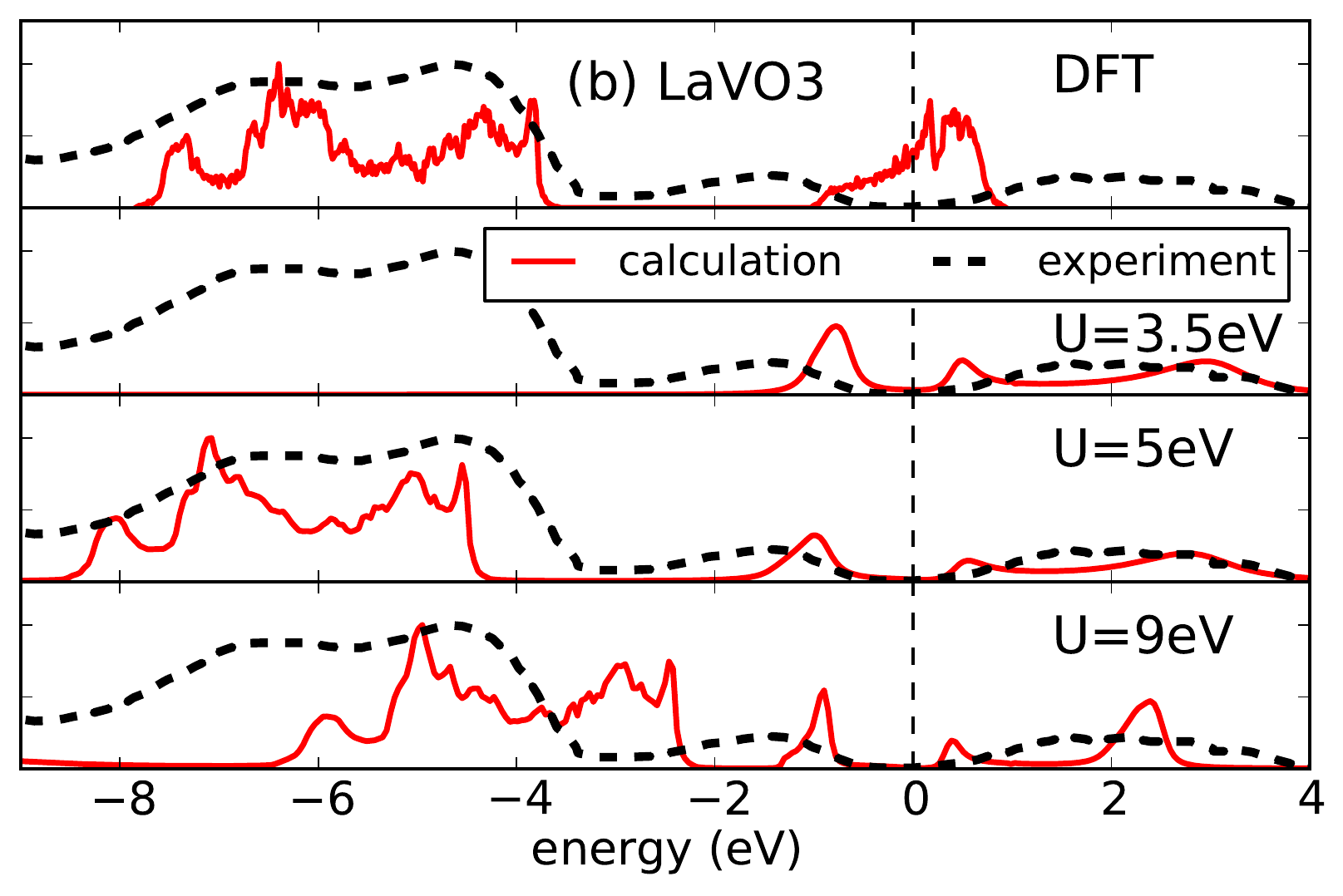}
      
    \caption{\label{fig:spec_dftnd_1}(Color online) Noninteracting density of states calculated using the Quantum Espresso implementation of the GGA approximation (with ``DFT'' label) and spectral functions $A(\omega)$ (solid curves) for LaTiO$_3$ (a - blue online) and LaVO$_3$ (b - red online) using the experimental lattice structure at $J=0.65eV$ for $U$-values indicated, with  $\Delta$ adjusted to match the insulating gap. The vertical dashed line marks the Fermi level. Black dashed curves give the experimentally determined oxygen density of states, reproduced from Refs.~\onlinecite{Maiti00,Imada98}. For LaTiO$_3$ the  $(U,N_d)$ pairs are  $(3.7,1.07)$, $(5.0,1.3)$ and $(9.0,1.5)$ with the band $N_d$ being $1.56$. For LaVO$_3$ we have $(3.7,2.05)$, $(5.0,2.30)$ and $(9.0,2.45)$ with the DFT value of $N_d$ being $2.55$. }
\end{figure}

To better understand the physics we present in  Fig.~\ref{fig:spec_dftnd_1} comparison of the calculated DFT and many-body density of states (DOS) for LaTiO$_3$ and LaVO$_3$ (solid lines)  along with experimental data\cite{Maiti00,Imada98} indicating the location of the oxygen bands (dashed lines). We see that the DFT ($U=0$) calculation places the oxygen bands about $1-1.5eV$  closer to the Fermi level than does experiment. For $U\ne0$ the $d$-level energy has been adjusted so that the DFT+DMFT calculation yields an insulating gap compatible with experiment ($0.2eV$ for LaTiO$_3$ and $1eV$ for LaVO$_3$).  For small or large values of interaction strength ($U\approx3.5eV$ and $U=9eV$), the calculated oxygen bands are either too far away (for $U\sim 3.5eV$) or too close to the Fermi level ($U=9eV$), but for $U\sim 5-6eV$, choosing the $\Delta$ ($N_d$) so the insulating gap is reproduced also yields a $p$-$d$ splitting in agreement with experiment.   As the cRPA calculation suggests that $U\sim 4eV$ for SrVO$_3$,\cite{Vaugier12} for more correlated materials such as LaTiO$_3$ or LaVO$_3$ the screening may be slightly weaker and $U$ value  slightly larger, so $U=5-6eV$ is a reasonable range. 

We now summarize the implications of our results. First, the phase diagrams presented in Fig.~\ref{fig:dmft_tilted_lto_lvo_1} show that the qualitative features of the metal-charge transfer insulator phase diagram previously reported \cite{Wang12} for the late transition metal oxides La$_2$CuO$_4$ and LaNiO$_3$ also apply to the early transition metal oxides. The phase diagram takes a simple form when presented in the $U-N_d$ plane, the phase boundary becomes vertical at large $U$, and for $N_d$ values similar to those predicted by DFT calculations increasing $U$ to arbitrarily large values does not drive a metal-insulator transition. In this sense, one may conclude that charge transfer physics is important in the titanate and vanadate materials, as well as in the cuprates and nickelates. Fig.~\ref{fig:dmft_tilted_lto_lvo_1} reveals, however, that physics in addition to $U$ and $N_d$ is important. In particular the offset between the DFT $N_d$ and that needed to drive a metal-insulator transition clearly depends on $d$-level filling. Also, as previously found by Pavarini and co-workers,\cite{Pavarini04} lattice structure (in particular the amplitude of the GdFeO$_3$ distortion) plays a crucial role in the metal-insulator transition in the $d^1$ (Ti) materials, and a noticeable but rather smaller role in the $d^2$ system. 

The spectra presented in Fig.~\ref{fig:spec_dftnd_1} provide further insights. We see that for a physically reasonable range of $U\sim 5-6$ eV, there exists a choice of double counting correction (i.e. a choice of $N_d$ ) which reproduces both the correct ground state (metallic for SrVO$_3$ and insulating for LaTiO$_3$ and LaVO$_3$) and the measured oxygen-band spectra. For this range of $U$-values, one can  see from the spectra in Fig.~\ref{fig:spec_dftnd_1} that the two materials have the same  difference between the DFT position of the oxygen-dominated bonding band and the measured position of this band (and indeed difference is the same as in SrVO$_3$). Thus to obtain agreement with experiment one must shift the $\Delta$ by the same amount in the two materials, although the needed shifts in $N_d$ are different. Finally, from Fig.~\ref{fig:spec_fullcharge_SVO_1} we see that the effect of full charge self-consistency and the standard FLL double-counting correction is to pin the $p$-$d$ energy difference to a definite value, slightly smaller than that provided by band theory even though $N_d$ changes with $U$.

Thus, we may conclude that for the early transition metal oxides the DFT+single-site DMFT procedure provides a good zeroth order picture of the electronic structure, provided that the $p$-$d$ splitting is correctly treated and the $U$ value is appropriately chosen. One may think about the problem of correcting the $p$-$d$ splitting in two ways. One is to regard it as arising from many-body physics in the $d$-level. In practical terms, the double counting correction acts to adjust  the position of the $d$-levels relative to other levels in the solid, so this approach corresponds to seeking a correct version of the double counting correction. Different versions have been explored in Refs.~\onlinecite{Karolak10,Haule13,Park13}. An alternative point of view is that  the focus should be shifted from the search for the correct double counting correction to the development of an electronic structure method which correctly positions the $p$ states relative to the $d$-states. One route to such a method might involve treating the $p$-$d$ portion of the Coulomb interaction in a beyond DFT approximation such as the GW method.\cite{Biermann03} Until such an electronic structure theory is developed or a rigorous theory of the double-counting correction is formulated, we suggest that attempts to model the physics of transition metal oxides should be based on a phenomenological double counting procedure chosen to reproduce the experimental $p$-$d$ energy difference and that care needs to be taken in fixing the value of the on site repulsion $U$.

{\it Acknowledgments:} We are very grateful to to Antoine Georges, Silke Biermann and Leonid Pourovskii for helpful conversations at various stages of this work and Michel Ferrero for providing an impurity solver. HTD and AJM  acknowledge support from DOE-ER046169 and CAM acknowledges funding from a DARPA Young Faculty Award, Grant Number D13AP00051. HTD acknowledges partial support from Vietnam Education Foundation (VEF). We acknowledge travel support from the Columbia-Sorbonne-Science-Po Ecole Polytechnique Alliance Program and thank Ecole Polytechnique (HTD and AJM) and J\"ulich Forschungszentrum (HTD) for hospitality while portions of this work were conducted. A portion of this research was conducted at the Center for Nanophase Materials Sciences, which is sponsored at Oak Ridge National Laboratory by the Scientific User Facilities Division, Office of Basic Energy Sciences, U.S. Department of Energy. This work also used the Extreme Science and Engineering Discovery Environment (XSEDE), which is supported by National Science Foundation grant number OCI-1053575.  We used the  CT-HYB solver\cite{PhysRevLett.97.076405} from the TRIQS project\cite{triqs_project} as well as one written by H. Hafermann, P. Werner and E. Gull\cite{Hafermann12} based on the  ALPS library.\cite{ALPS1.0,ALPS2.0}

\end{document}